\documentclass[12pt,doublespace,onecolumn]{IEEEtran}
\usepackage{cite}
\usepackage{amsmath}
\usepackage{amssymb}
\usepackage{graphicx}
\usepackage{epsfig}
\title{AutoAmp~:~An Open-Source Analog Amplifier Design Tool~-~For Classroom and Lab Purposes}
\author{Om Prasad Patri ~\and~ K~Sanmukh Rao}

\date{}
\begin{document}
\maketitle
\begin{center}
$\{$o.patri,~kuppannagari$\}$\textcircled{a}iitg.ernet.in
\end{center}

\begin{center}
 Supervised by : Dr.~Amit K. Mishra, Assistant Professor, Department of ECE, IIT Guwahati
\end{center}

\begin{center}
 Department of Computer Science and Engineering,
 \\Indian Institute of Technology Guwahati
\end{center}

\begin{abstract}
This correspondence presents an open-source tool \emph{AutoAmp} developed at the Indian Institute of Technology,~Guwahati. It is
available at \texttt{\underline{http://sourceforge.net/projects/autoamp-iitg/}} 
This tool helps the user to design different types of electronic amplifiers, using solid state devices, for a given specification. 
It can handle several types of designs namely common-emitter BJT amplifier (single and two-stage), operational amplifiers (inverting and non-inverting) and power amplifier. 
 Not only does it design the amplifier, it also simulates the designed amplifier using SPICE simulator and displays the performance curves. This tool is deemed to prove invaluable in undergraduate teaching and labs. Especially in electronics-design related laboratories, the student need not design the amplifiers which are mostly the heart of many electronic designs. 

\end{abstract}

\section{Introduction}
Analog amplifiers are the building blocks of many electronic circuits. Different types of analog electronic amplifiers are commonly used in radio and television transmitters and receivers,~high fidelity (hi-fi) stereo equipment, micro computers and other electronic equipment. 
Transistor amplifiers are among the most commonly used kinds of amplifiers. Most common active devices used in transistor based amplifiers are bipolar junction transistors~(BJTs) and metal oxide semiconductor field-effect transistors~(MOSFETs),~with BJTs being a preferred choice for lab level circuit-design. BJT based amplifiers find use in audio amplifiers in a home stereo or public address system,~RF high power generation for semiconductor equipment,~RF and microwave applications such as radio transmitters. An operational amplifier(\emph{op-amp}) is an amplifier circuit with very high open loop gain and differential inputs which employs external negative feedback for control of its transfer function and gain. These attributes form the basis for op-amp applications in integrated circuits and its extensive study and use in experimental circuits.
 
We have developed a simple open-source amplifier design tool,~named AutoAmp,~\cite{autoamp}~for the following types of amplifiers, given some design specifications:
\begin{enumerate}
 \item Single Stage BJT CE Amplifier
 \item Two-Stage BJT CE Amplifier
 \item Operational Amplifier:~Inverting/Non-Inverting
 \item Op-Amp Difference Amplifier
 \item Class-A Power Amplifier
\end{enumerate}
For each type of design the software requires minimum design specifications assuming the lab working environment. The assumptions can however be modified in the source code as per requirements. Net voltage gain is the key design parameter that the software uses across most amplifier types. However, few designs require more specific information like the maximum available resistance in the case of operational amplifier design;  power transmitted to load, $V_{CC}$ and load resistance in the case of power amplifier.
Given the choice of type followed by minimum design specifications, the software generates a netlist file to be  opened in LTSpice\texttrademark{}~\cite{LTSpice}~for necessary circuit analysis. AutoAmp is open-source and can be run on both Windows and Linux-based systems. The necessary adaptations for the OS are mentioned in the AutoAmp website~\cite{autoamp}. Autoamp is available for free (along with the source code) at \cite{autoamp}.

There are not many free tools which automatically design basic amplifier circuits given the design specifications. It is expected that industries may be maintaining customized circuit design tools to solve their purpose. 
However use of such commercial tools for academic purposes is likely to be prohibitively expensive. Basic amplifier-circuit design along with its analysis is required in any complex circuit in electronics. Thus availability of such a tool will be a boon for teachers and students alike.    

Among the existing tools for amplifier circuit design, tools for operational amplifiers are available, including online tools, like \cite{onlinetool},~however, there is no open-source design tool available for designing amplifier circuits with BJT amplifiers and class-A power amplifiers. Most importantly, none of the available tools have a provision for the analysis of the circuit generated. We aim to provide an interface where the user can get the design of the circuit in LTSpice\texttrademark{} after providing certain design specifications. Such a tool will be very useful in classrooms and for other non-industrial purposes where such circuit design is warranted.

The existing design tools are pretty complicated (especially for classroom purposes), difficult to use, expensive, not open-source (user cannot change the source to suit his own purpose) and lack a Spice or similar interface. Moreover, design tools for BJTs, power amplifiers are very hard to find. The commercial tools come with a whole package of electronic design automation tools with lot of circuit-options, which makes them complicated. For learning or teaching a course in analog electronics, only a few numbers of these circuits are required. Further, addition or deletion of components and changing the source according to individual requirements can not be done.

Our design tool tries to overcome most of these problems. It is a simple and user-freindly tool.  AutoAmp is easy to operate, takes minimum input and generates an LTSpice\texttrademark{} netlist which can be used to design the circuit in LTSpice\texttrademark{} directly. Being open-source, customized changes can be easily made to the source code to give the desired results; components can be easily added or removed by writing some extra functions in the source code.

Section \ref{proposal} describes our design approach in detail including the software design methodology and the circuit design approach. Section \ref{demo} shows our demo/experimentation results and includes screenshots from the working of the tool. The elaborate theoretical and mathematical analysis for each type of amplifier can be found in the Appendix \ref{appsingle} . Section \ref{conclusion} sums up the proposal in the conclusion and talks about possible future work related to the tool. 

\section{Design approach\label{proposal}}
\begin{figure}[h]
 \centering
 \includegraphics[width=0.6\textwidth,angle=0]{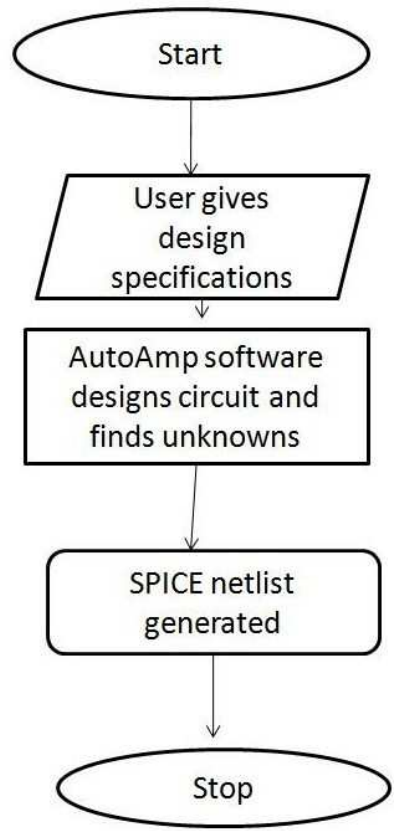}
 \caption{Blackbox Diagram for AutoAmp \label{flowchart}}
\end{figure}

This section describes our design approach of the tool in detail. A blackbox representation of the tool is given by Fig.~\ref{flowchart}.

\subsection{Software Design Methodology}
The tool is a command line software designed in C++ programming language. The program has a class named autoAmp which consists of various functions for computation of the amplifiers' components and one function for printing in the file. A struct data type is defined to store all the computed values and is finally used to create the output file. The user is asked for the name of the input file, to select an amplifier of her choice in a menu based environment and finally to enter voltage gain and other parameters based on the type of amplifier chosen. Based on user's choice the respective functions are called which compute the values of components and store then into the struct defined. Now another function uses this struct to create the netlist of the respective type of amplifier into the file specified by the user in the beginning.

\subsection{Circuit Design Approach}
\subsubsection{Single Stage BJT CE Amplifier}
We have a designed a small-signal voltage amplifier operating in the audio frequency range. We have used an n-p-n transistor, namely, 2N2222. Two port h-parameters are used for circuit analysis. Maximum, minimum, and typical values as required,~of the h-parameters are obtained from the transistor datasheet~\cite{datasheet}. One method for obtaining the hybrid parameters of the BJT amplifier is given by Al-Zobi et al~\cite{ieeepaper}.~These values along with the other known values are used by the software to get optimized values of the circuit components, i.e., resistances and capacitors. The detailed theoretical and mathematical analysis for this part can be found in the appendix~\ref{appsingle}.
\subsubsection{Two Stage BJT CE Amplifier}
This design consists of two CE amplifier stages in cascade.~Two amplifying stages thus give us a higher overall gain. The design methodology remains the same as in single stage CE amplifier but is applicable over two stages in this case. The detailed theoretical and mathematical analysis for this part can be found in the appendix~\ref{appdouble}.
\subsubsection{Operating Amplifiers}
This is the simplest ampifier designing strategy in which we are using the inverting and non-inverting configurations of the ideal Universal OpAmp2 (as given in LTSpice\texttrademark). The detailed analysis is given by Sedra and Smith~\cite{sedrasmith}. Assuming an ideal op amp with infinite open-loop gain, $R_{in}$ = $R_1$. Now to avoid the loss of signal strength, voltage amplifiers are required to have high input resistance. In the case of the inverting op-amp confguration we are studying, to make $R_{in}$ high we should select a high value for $R$. However, if the required gain $\frac{R_2}{R_1}$ is also high then $R$ could become impratically large (e.g.~greater than a few megaohms). Hence in our design we use a different feedback mechanism by which the circuit is able to realize a large voltage gain without using large resistances in the feedback path. The details of the design can be found in the Appendix. Standard circuit design can be used for the non-inverting input configuration as the input resistance is infinity as desired. The detailed theoretical and mathematical analysis for this part can be found in the appendix~\ref{appopamp}.
\subsubsection{Op-amp Difference Amplifier}
In this design, we implement a difference amplifier again using Universal OpAmp2 that responds to the difference between the two signals applied at its input and ideally rejects signals that are common to the two inputs. The circuit design uses four resistances.~The detailed theoretical and mathematical analysis for this part (along with the circuit diagram) can be found in the appendix~\ref{appopampdiff}.
\subsubsection{Class-A Power Amplifier} 
Here we have designed a power amplifier by simply implementing a CE BJT amplifier with a high power output stage for which a transformer is used. The detailed theoretical and mathematical analysis for this part can be found in the appendix~\ref{apppower}.

\section{Demonstration\label{demo}}
\begin{figure}[htbp]
 \centering
 \includegraphics[width=1\textwidth]{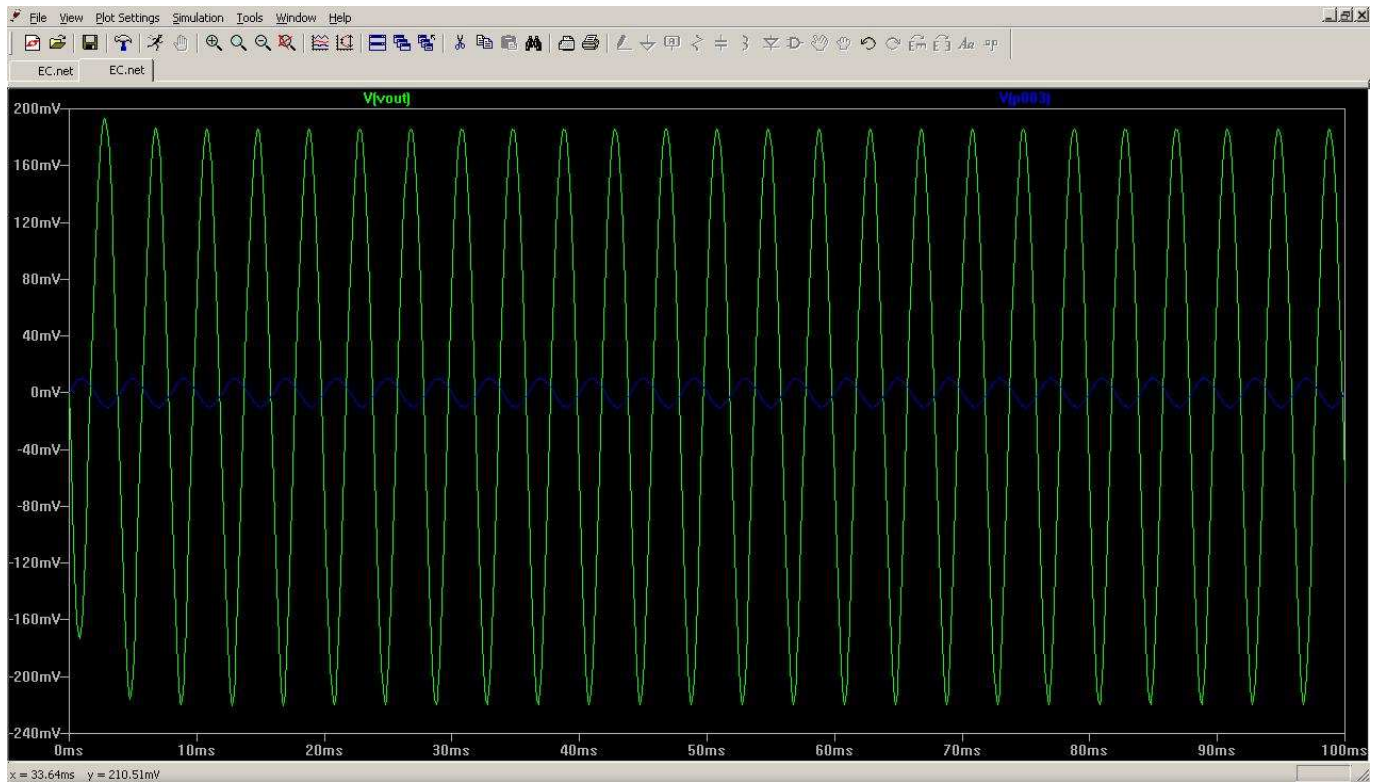}
 \caption{$V_{in}$ v/s $V_{out}$ in BJT Single stage Amplifier\label{ss_bjt1}}
\end{figure}

\begin{figure}[htbp]
 \centering
 \includegraphics[width=1\textwidth]{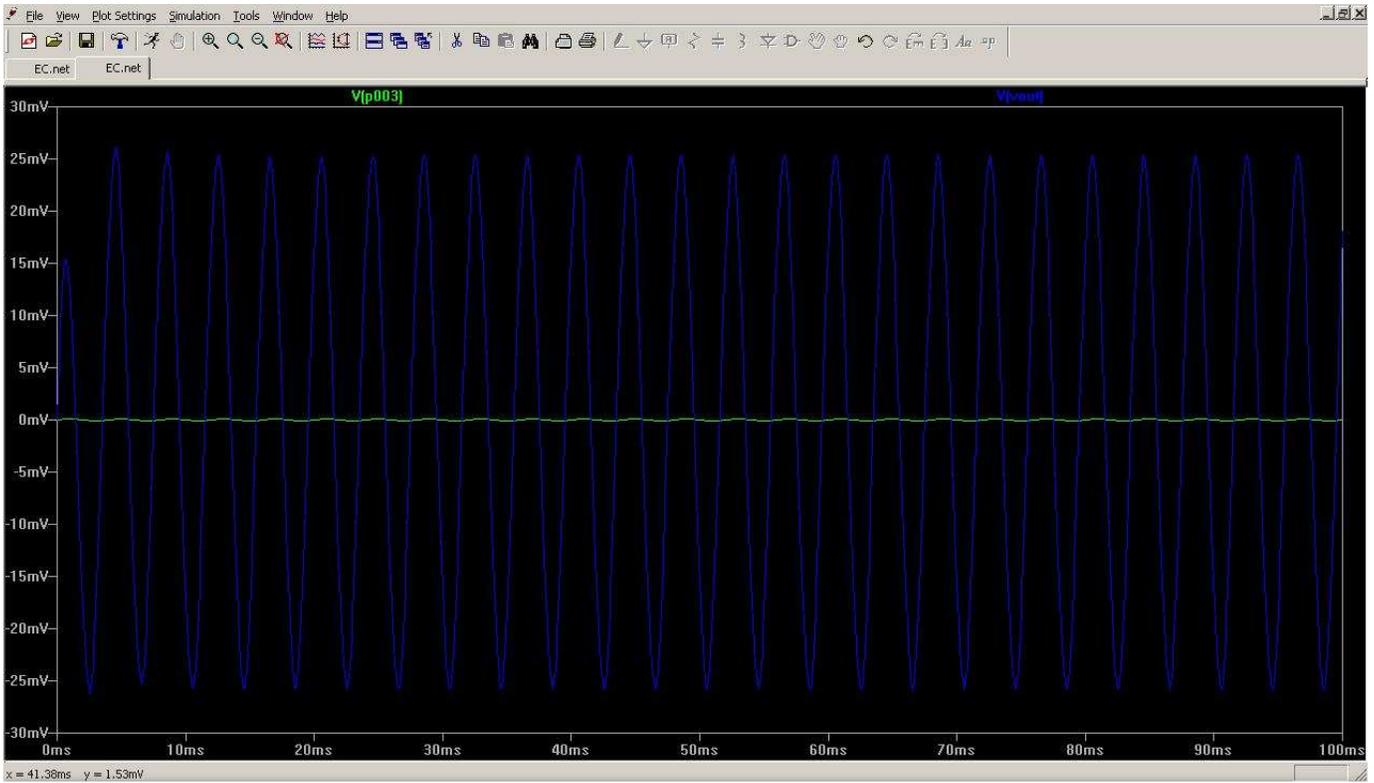}
 \caption{$V_{in}$ v/s $V_{out}$ in BJT two-stage Amplifier\label{ss_bjt2}}
\end{figure}

\begin{figure}[tbp]
 \centering
 \includegraphics[width=1\textwidth]{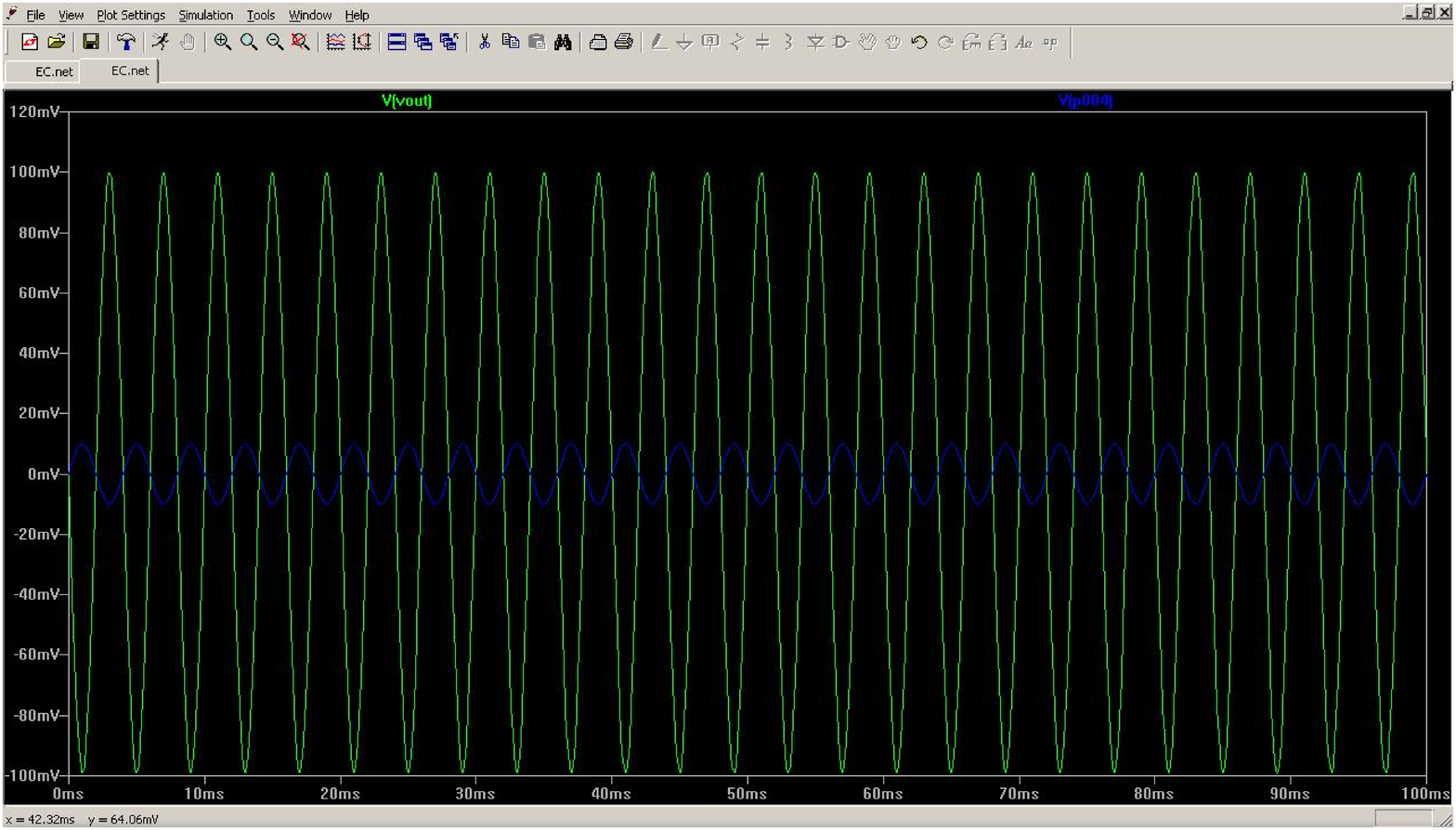}
 \caption{$V_{in}$ v/s $V_{out}$ in Inverting Op-amp\label{ss_inv}}
\end{figure}

\begin{figure}[tbp]
 \centering
 \includegraphics[width=1\textwidth]{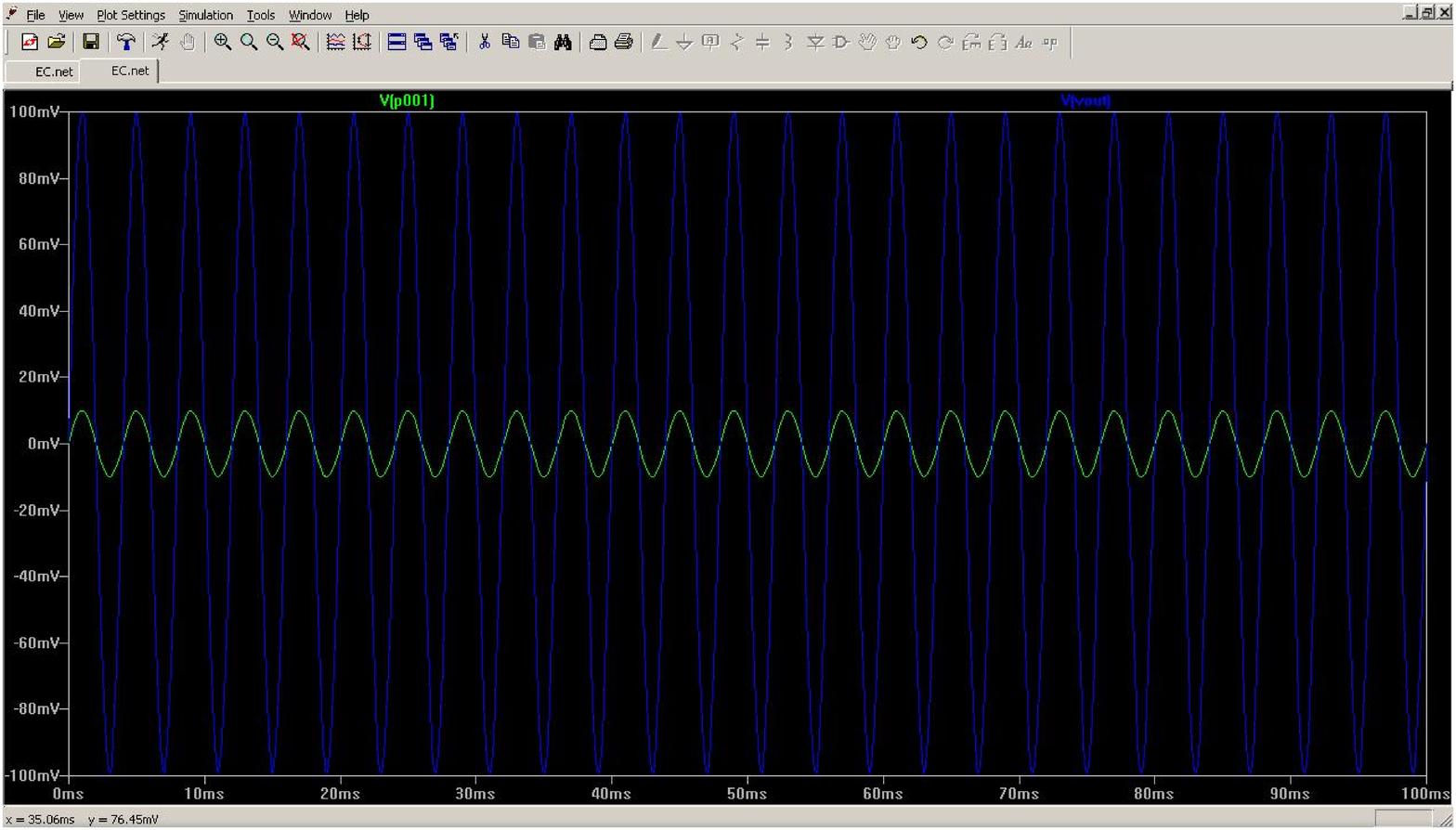}
 \caption{$V_{in}$ v/s $V_{out}$ in Non-Inverting Op-amp\label{ss_noninv}}
\end{figure}

\begin{figure}[tbp]
 \centering
 \includegraphics[width=1\textwidth]{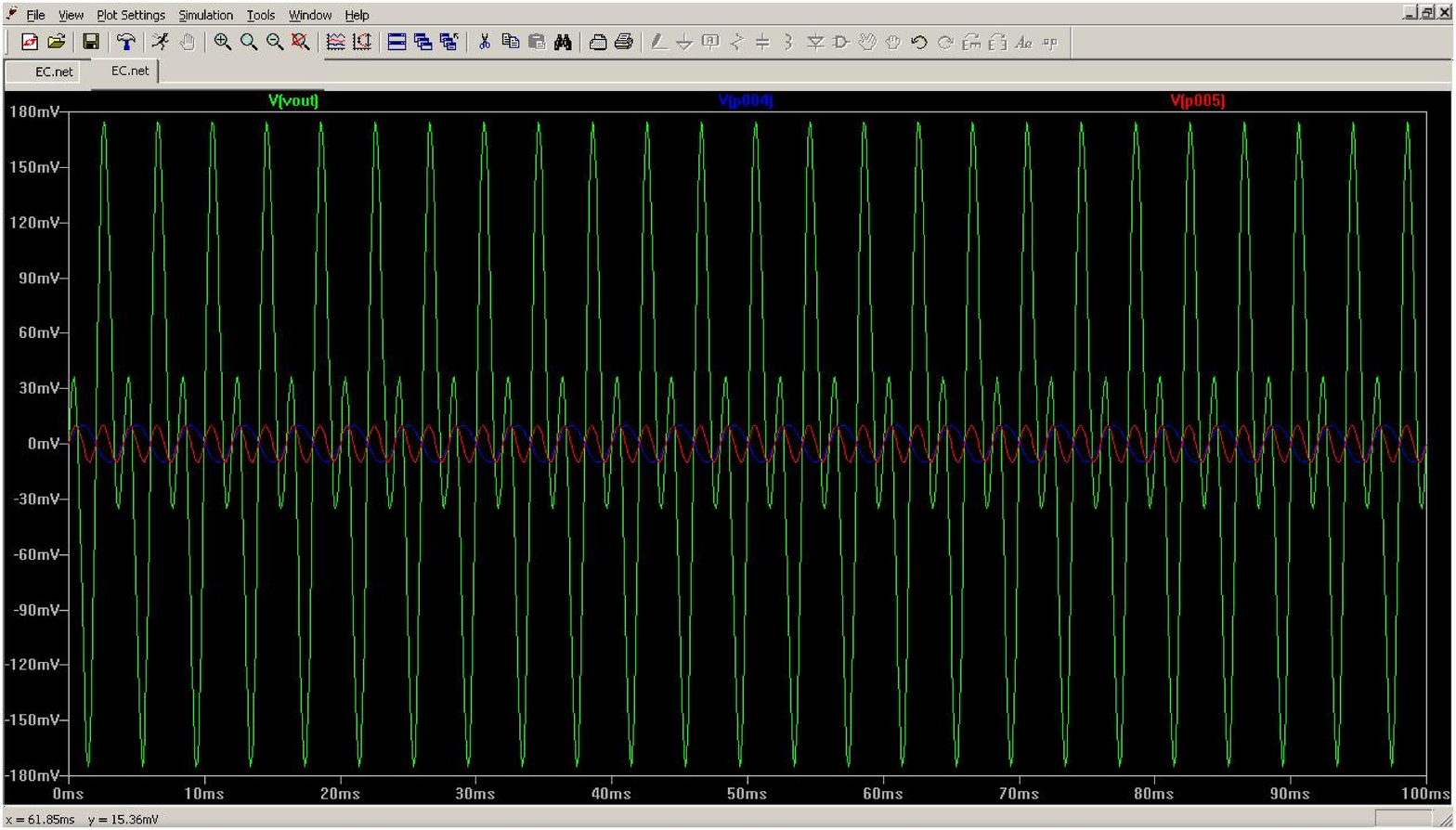}
 \caption{$V_{in}$ v/s $V_{out}$ in Op-amp based Difference Amplifier\label{ss_diff}}
\end{figure}

Here are some screenshots which are obtained after the netlist file created by AutoAmp is input to LTSpice\texttrademark. We simulated the netlist files generated by AutoAmp in LTSpice. The input-output voltage graphs for sinusoidal input voltages are presented in the figures \ref{ss_bjt1} - \ref{ss_diff}. Figure \ref{ss_bjt1} shows a voltage gain of $20$ for an input of $20 mV$ peak-to-peak in single-stage BJT amplifier. Figure \ref{ss_bjt2} shows a total voltage gain of $100$ for an input of $0.5 mV$ peak-to-peak in a two-stage BJT amplifier. Figure \ref{ss_inv} and Figure \ref{ss_noninv} show a voltage gain of $10$ for an input of $20 mV$ peak-to-peak in inverting and non-inverting operational amplifier. Figure \ref{ss_diff} shows the graph of voltage gain for difference operational amplifier.
\\

\section{Conclusion\label{conclusion}}
This paper presents a simple open-source design tool using C++ which can help in designing and analyzing an amplifier given some design specifications. 
The software is able to demonstrate five design spec-combinations. 
The complete package is deemed to be of utility to circuit designers and instructors, especially in a technical college environment.

\section*{Acknowledgements}
The authors are grateful to Dr. Amit Kumar Mishra, Assistant Professor, Department of Electronics and Communication Engineering, IIT Guwahati for
 supervising this work. Also, they thank their colleague Mr. Nitin Dua, Department of Computer Science and Engineering, IIT Guwahati for his 
help with the theoretical analysis of the various amplifier circuits.

\bibliographystyle{IEEEtran}
\bibliography{bibliography}


\appendix
\subsection{Design of single stage CE BJT Amplifier \label{appsingle}} 
We show the design of the common-emitter single stage BJT audio frequency amplifier.~$A_V$ and $V_0$ is input by the user. We use voltage divider circuit as it provides Q point independent of temperature and beta. We use H-parameters of the CE BJT to analyse the circuit. The H parameters are $h_{fe}$,~$h_{oe}$,~$h_{ie}$ and $h_{re}$,~where $h_{fe}$ is the current gain, $h_{ie}$ is the input impedance and $h_{oe}$ is the output admittance.\\
\begin{figure}[h]
 \centering
 \includegraphics[width=0.5\textwidth]{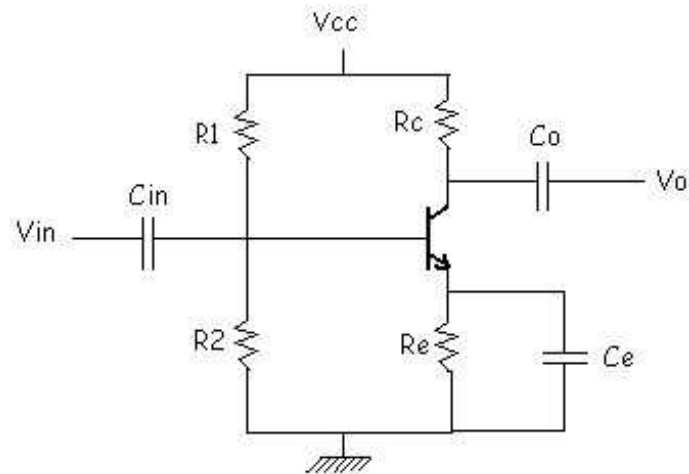}
 \caption{Single stage BJT CE amplifier}
\end{figure}

\emph{Selection of $R_C$}\\
\begin{equation}
R_{L'} = R_{C} \parallel R_{L}
\end{equation}
If $R_L$ not given, $R_{L'}$ = $R_{C}$\\
$R_L$ = load resistance connected between $V_0$ and ground.\\
\begin{equation}
A_V = \frac{h_{fe} \ast R_{L'}}{h_{ie} + (h \ast R_C)}
\end{equation}
Where \begin{equation}
h = (h_{ie} \ast h_{oe}) - (h_{fe} \ast h_{re})\\
\end{equation}
We got $h_{fe}$,~$h_{re}$,~$h_{ie}$,~$h_{oe}$~from data sheet \cite{datasheet}.\\
\begin{equation}
h_{fe} = \beta,~h_{fe,max} = \beta_{max},~h_{fe,min} = \beta_{min}
\end{equation}
From the above equations we calculated $R_{L'}$ and $R_C$. Select higher std value for $R_C$ to increase voltage
gain if min voltage gain is specified or nothing is specified. If max voltage gain is specified use lower std value. If some specific voltage gain is specified, use nearest std val.\\
\\
\emph{Selection of operating point ($V_{CEQ}$,~$I_{CQ}$)}\\
If $V_{CC}$ is given,
\begin{equation}V_{CEQ} = \frac{V_{CC}}{2}\end{equation}
If V$_{CC}$ is not given,
\begin{equation}V_{CEQ} = 1.5~\ast~(V_{0,peak} + V_{CE,sat})\end{equation}
Then,
\begin{equation}
I_{C,peak} = \frac{V_{0,peak}}{R_{L'}} 
\end{equation}
\begin{equation}
I_{CQ} = I_{C,peak} + I_{C,min}
\end{equation}
Assume $I_{C,min}$ = $0$ or $0.005 mA$\\
\\
\emph{Selection of $R_e$}\\
If $V_{CC}$ is not given,~assume \begin{equation*}V_{re} = 1\end{equation*}
If $V_{CC}$ is given,\begin{equation*}V_{re} = 10\%~of~V_{CC}\end{equation*}
For either case,
\begin{equation}
 R_e = \frac{V_{re}}{I_{CQ}}
\end{equation}
Select lower std value of $R_e$ so that voltage drop across $R_e$ is less which increases the voltage swing of o/p.\\
\\
\emph{Selection of $V_{CC}$}\\
If $V_{CC}$ is not given,\begin{equation} V_{CC} = V_{CEQ} + I_{CQ}\ast(R_C + R_e)\end{equation}
Assume higher std val (typically 9,12,15,18).\\
\\
\emph{Selection of $R_1$ and $R_2$}\\
If stability factor is not given,~assume $s$ = 8\\
\begin{equation}
s = \frac{1 + h_{fe,max}}{1 + \frac{h_{fe,max}~\ast~R_e}{R_b + R_e}}
\end{equation}
$R_b$ is found and it is not standardized.\\
\begin{equation}
V_{r2} = V_{be} + V_{re} \qquad V_{r1} = V_{CC} - V_{r2}
\end{equation}
Assume $V_{be}$ = 0.7~V (for Si)\\
\begin{equation}
\frac{R_1}{R_2} = \frac{V_{r1}}{V_{r2}} \label{1stageeqn}
\end{equation}
We get $R_1$ in terms of $R_2$. Substitute in $R_b$.
\begin{equation}
R_b = R_1 \parallel R_2 = \frac{R_1 \ast R_2}{R_1 + R_2}
\end{equation}
$R_2$ is found. Select lower standard value to make circuit independent of $\beta$. Substitute in \eqref{1stageeqn} to find $R_1$. We should select higher standard value so that circuit draws minimum current from supply.\\
\\
\emph{Selection of coupling capacitors}\\
Select higher standard value for all capacitors.\\
\textit{Selection of $C_E$}\\
\begin{equation}
X_{CE} = \frac{R_e}{10}
\end{equation}
\begin{equation}
C_E = \frac{1}{2~\ast \pi~\ast f_L~\ast X_{CE}}
\end{equation}
where $f_L$ = lower cutoff frequency. Assume $f_L$ = 20 Hz.\\
\textit{Selection of $C_B$}\\
If $R_S$ (Source resistance) is not specified, assume $R_S = 0$\\
\begin{equation} X_{CB} = R_S + R_b \parallel h_{ie} \end{equation}
\begin{equation} C_B = \frac{1}{2~\ast \pi~\ast f_L~\ast X_{CB}} \end{equation}
\textit{Selection of $C_C$}
\begin{equation} R_b = R_1 \parallel R_2 \end{equation}
\begin{equation} X_{CC} = R_C + R_L \end{equation}
If $R_L$ (load resistance) is not specified, then we assume amplifier is connected to a similar next stage.\\
Hence, \begin{equation} R_L = R_b \parallel h_{ie} \end{equation}
\begin{equation} C_C = \frac{1}{2~\ast~\pi~\ast~f_L\ast~X_{CC}} \end{equation}

\subsection{Design of two-stage BJT CE Amplifier\label{appdouble}}
Value of $A_V$ is input by the user.\\
Overall voltage gain, \begin{equation} A_V = A_{v1} \ast A_{v2} \end{equation}
\begin{equation} \frac{A_{v1}}{A_{v2}} = \frac{R_{C1}}{R_{C2}} \tag{Assumed} \end{equation}
$A_{v1}$ and $A_{v2}$ can be found from the above equations.\\
\begin{figure}[ht]
 \centering
 \includegraphics[width=0.95\textwidth,angle=0]{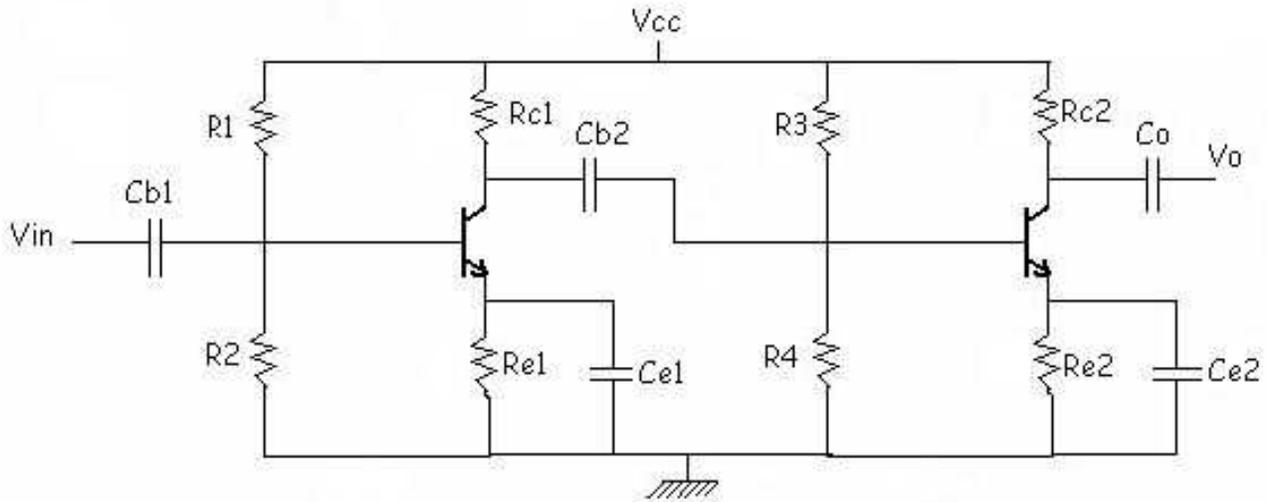}
 \caption{Two-stage BJT CE amplifier}
\end{figure}
\subsubsection{Part 1:~Design of second stage}
\emph{Calculation of $R_L$:}\\
\begin{equation}
R_L = \frac{V_{0,peak}}{I_{0,peak}}
\end{equation}
Not necessary if $R_L$ is not given or $V_{0,peak},~I_{0,peak}$ is not given.\\
\\
\emph{Selection of $R_{C2}$:}\\
\begin{equation}
A_{v2} = \frac{h_{fe} \ast R_{L2'}}{h_{ie}}
\end{equation}
\begin{equation}
R_{L'} = R_{C2} \parallel R_L
\end{equation}
or, if $R_L$ is not given, then,
\begin{equation}
R_{L'} = R_{C2}
\end{equation}
Hence, we calculate $R_L'$ and $R_{C2}$ \\
\\
\emph{Selection of $V_{CEQ}$:}\\
\textit{Case 1:~$V_{CC}$ is given}\\
\begin{equation}
V_{CEQ,2} = 1.5~\ast~(V_{0,peak} + V_{CE,saturation})
\end{equation}
If $V_0$ is not given,~then\\
\begin{equation}
V_{CEQ} = \frac{V_{CC}}{2}
\end{equation}
\begin{equation}
V_{re2} = 10~\text{to}~20~\%~\text{of} V_{CC}
\end{equation}
\begin{equation}
V_{rc2} = V_{CC} - V_{CEQ,2} - V_{re2}
\end{equation}
\begin{equation}
I_{CQ,2} = \frac{V_{rc2}}{R_{C2}}
\end{equation}
\begin{equation}
R_{e2} = \frac{V_{re2}}{I_{CQ,2}}
\end{equation}
Select lower standard value so that drop across $R_e$ is less which increases gain of the output.\\
\textit{Case 2:~$V_CC$ is not given}\\
\begin{equation}
V_{CEQ,2} = 1.5~\ast~(V_{0,peak} + V_{CE,saturation})
\end{equation}
\begin{equation}
I_{C2,peak} = \frac{V_{0,peak}}{R_{L2}}
\end{equation}
Assume $V_{re2} = 2$ V
\begin{equation}
V_{CC} = V_{CEQ} + I_{CQ}~\ast~(R_{C2} + R_{e2})
\end{equation}
Select higher std value.\\
\\
\emph{Selection of $R_3$, $R_4$}\\
If stability factor is not given, assume $s$ = 8\\
\begin{equation}
s = \frac{h_{fe} + 1}{1 + \frac{h_{fe}~\ast~R_{e2}}{R_{b2} + R_{e2}}}
\end{equation}
$R_b$ is found.\\
\begin{equation}
V_{r4} = V_{be} + V_{re}
\end{equation}
\begin{equation}
V_{r3} = V_{CC} - V_{r2}
\end{equation}
Assume $V_{be}$ = 0.7V (for Si)\\
\begin{equation}
\frac{R_3}{R_4} = \frac{V_{r3}}{V_{r4}} \label{2stageeqn1}
\end{equation}
Get $R_3$ in terms of $R_4$ and substitute in $R_{b2}$
\begin{equation}
R_{b2} = R_3 \parallel R_4 = \frac{R_3~\ast~R_4}{R_3 + R_4}
\end{equation}
Find $R_4$. Select lower standard value to make circuit independent of beta. Substitute in \eqref{2stageeqn1} to find $R_3$. Select higher standard value so that circuit draws minimum current from supply.\\
\subsubsection{Part 2:~Design of first stage}
\begin{equation}
A_{v1} = \frac{A_V}{A_{v2}}
\end{equation}
\emph{Selection of $R_{C1}$}\\
\begin{equation}
A_{v1} = \frac{h_{fe}~\ast~R_{L1'}}{h_{ie}}
\end{equation}
\begin{equation}
R_{L'} = R_{C1} \parallel R_{b2} \parallel h_{ie}
\end{equation}
$R_{L'}$ and $R_{C1}$ is calculated.\\
Let $V_{CEQ,1} = V_{CEQ,2}, ~V_{rc1} = V_{rc2},~V_{re1} = V_{re2},~I_{CQ,1} = \frac{V_{rc1}}{R_{C1}}~and~R_{e1} = \frac{V_{re1}}{I_{CQ,1}}$\\
\\
\emph{Selection of $R_1$, $R_2$}
\begin{equation}
s = \frac{1 + h_{fe,max}}{1 + \frac{h_{fe,max}~\ast~R_e}{R_b + R_e}}
\end{equation}
$R_b$ is found.\\
\begin{equation}
V_{r2} = V_{be} + V_{re}
\end{equation}
\begin{equation}
V_{r1} = V_{CC} - V_{r2}
\end{equation}
Assume $V_{be}$ = 0.7V\\
\begin{equation}
\frac{R_1}{R_2} = \frac{V_{r1}}{V_{r2}} \label{2stageeqn2}
\end{equation}
We get $R_1$ in terms of $R_2$ and substitute in $R_b$.
\begin{equation}
R_b = R_1 \parallel R_2 = \frac{R_1~\ast~R_2}{R_1 + R_2}
\end{equation}
Find $R_2$.~We select lower standard value to make circuit independent of $\beta$.~Substitute in \eqref{2stageeqn2} to find $R_1$.~Select higher standard value so that circuit draws minimum current from supply.\\
\\
\emph{Selection of coupling capacitors:}
Select higher standard value for all capacitors.\\
\textit{Selection of $C_{E1}$:}
\begin{equation}
X_{CE1} = \frac{R_{e1}}{10}
\end{equation}
\begin{equation}
C_{E1} = \frac{1}{2~\ast~\pi~\ast~f_L~\ast~X_{CE1}}
\end{equation}
$f_L$ = lower cutoff frequency.~We assume $f_L$ = 20 Hz.\\
\textit{Selection of $C_{E2}$:}\\
\begin{equation}
X_{CE2} = \frac{R_{e2}}{10}
\end{equation}
\begin{equation}
C_{E2} = \frac{1}{2~\ast~\pi~\ast~f_L~\ast~X_{CE2}}
\end{equation}
\textit{Selection of $C_{B1}$:}
\begin{equation}
R_b = R_1 \parallel R_2
\end{equation}
If $R_S$ (Source resistance) is not specified, assume $R_S$ = 0\\
\begin{equation}
X_{CB1} = R_b \parallel h_{ie}
\end{equation}
\begin{equation}
C_{B1} = \frac{1}{2~\ast~\pi~\ast~f_L~\ast~X_{CB}}
\end{equation}
\textit{Selection of $C_{B2}$:}\\
\begin{equation}
R_{b2} = R_3 \parallel R_4
\end{equation}
\begin{equation}
X_{CB2} = R_{C1} + R_b \parallel h_{ie}
\end{equation}
\begin{equation}
C_{B2} = \frac{1}{2~\ast~\pi~\ast~f_L~\ast~X_{CB}}
\end{equation}
\textit{Selection of $C_0$:}\\
\begin{equation}
R_{b2} = R_3 \parallel R_4
\end{equation}
\begin{equation}
X_{CC} = R_C + R_L
\end{equation}
If $R_L$~(load resistance) is not specified,~then assume amplifier is connected to a similar next stage.~Hence,
\begin{equation}
R_L = R_{b1} \parallel h_{ie}
\end{equation}
\begin{equation}
C_C = \frac{1}{2~\ast~\pi~\ast~f_L~\ast~X_{CC}}
\end{equation}

\subsection{Design of Operational Amplifiers based Amplifiers\label{appopamp}}
Depending upon whether the value of $A_V$ input by the user is positive or negative,~the circuit is interpreted to be of the configuration non-inverting or inverting,~respectively.\\
\subsubsection{Non-Inverting Amplifier}
\begin{figure}[htp]
 \centering
 \includegraphics[width=0.5\textwidth]{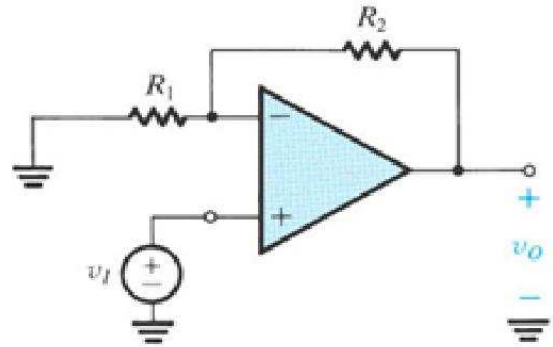}
 \caption{Non-Inverting Operational Amplifier \cite{sedrasmithstud}}
\end{figure}

The user inputs the gain $A_V$ and the resistances $R_1$ and $R_2$ are found by the following formula~:\\
\begin{equation}
A_V = 1 + \frac{R_2}{R_1}
\end{equation}

\subsubsection{Inverting Amplifier}
\begin{figure}[htp]
 \centering
 \includegraphics[width=0.5\textwidth]{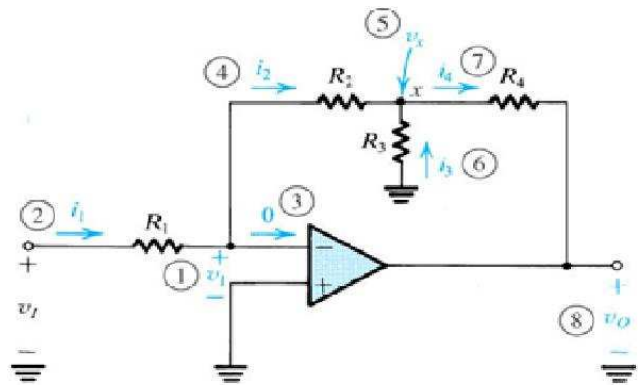}
 \caption{Inverting Operational Amplifier \cite{sedrasmithstud}}
\end{figure}

At the inverting terminal of the op-amp,~the voltage is\\
\begin{equation}
V_1 = -\frac{v_0}{A} = -\frac{v_0}{\infty} = 0
\end{equation}
Here we have assumed that the circuit is producing a finite output voltage $v_0$.\\
\begin{equation}
i_1 = \frac{v_I - v_1}{R_1} = \frac{v_I - 0}{R_1} = \frac{v_I}{R_1}
\end{equation}
Since zero current flows into the inverting input terminal,~all of $i_1$ will flow through $R_2$, and thus~:\\
\begin{equation}
i_2 = i_1 = \frac{v_{in}}{R_1}
\end{equation}
Voltage at $x$~:
\begin{equation}
V_x = v_1 - i_2~\ast~R_2 = 0 - \frac{v_I}{R_1}~\ast~R_2 = -\frac{R_2}{R_1}~\ast~v_I
\end{equation}
This in turn enables us to find $i_3$~:
\begin{equation}
i_3 = \frac{0 - V_x}{R_3} = \frac{R_2}{R_1~\ast~R_3}~\ast~v_I
\end{equation}
Next,~a node equation at $x$ yields $i_4$~:
\begin{equation}
i_4 = i_2 + i_3 = \frac{v_I}{R_1} + \frac{R_2}{R_1~\ast~R_3}~\ast~v_I
\end{equation}
Finally,~we can determine $v_0$ from
\begin{equation}
v_0 = V_x - i_4~\ast~R_4 = (-\frac{R_2}{R_1})~\ast~v_I - (\frac{v_I}{R_1} + \frac{R_2}{R_1~\ast~R_3}~\ast~v_I)~\ast~R_4
\end{equation}
Thus,~ the voltage gain is given by~:
\begin{equation}
\frac{v_0}{v_I} = \frac{R_2}{R_1} + \frac{R_4}{R_1}~\ast~(1 + \frac{R_2}{R_3})
\end{equation}
which can be written in the form
\begin{equation}
\frac{v_0}{v_I} = -\frac{R_2}{R_1}~\ast~(1 + \frac{R_4}{R_2} + \frac{R_4}{R_3})
\end{equation}

\subsection{Design of Op-Amp based Difference Amplifier \label{appopampdiff}}
User inputs $A_d$.\\
In ideal case,
\begin{equation}
A_{CM} = 0 \qquad \text{and} \qquad A_d = \frac{R_2}{R_1}
\end{equation}

\subsection{Design of Power Amplifier Class A\label{apppower}}
\begin{figure}[htp]
 \centering
 \includegraphics[width=0.6\textwidth]{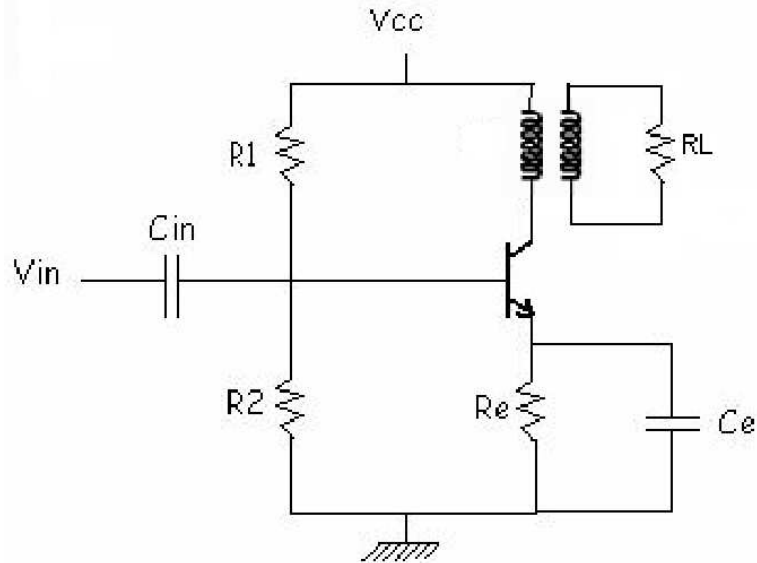}
 \caption{Power Amplifier Class A}
\end{figure}

\emph{Selection of operating point~:}
\begin{equation}
V_{re} = \frac{V_CC}{10}
\end{equation}
\begin{equation}
V_{CEQ} = V_{CC} - V_{re}
\end{equation}
\begin{equation}
V_{CE,peak} = V_{CEQ} - V_{CE,sat}
\end{equation}
\begin{equation}
I_{C,peak} = \frac{2~\ast~P_{L'}}{V_{CE,peak}}
\end{equation}
\begin{equation}
I_{CQ} = I_{C,peak} + I_{C,min}
\end{equation}
Assume $I_{C,min}$ = 0.\\
Hence, we calculate $I_{CQ}$\\
\\
\emph{Selection of $R_e$ and $C_e$~:}
\begin{equation}
R_e = \frac{V_{re}}{I_{CQ}}
\end{equation}
\begin{equation}
P_{re} = \frac{V_{re}^2}{R_e}
\end{equation}
\begin{equation}
C_e = \frac{1}{2~\ast~\pi~\ast~f_L~\ast~R_L}
\end{equation}
We assume $f_L$ = 50 Hz. Since $C_e$ is very high,~we leave $R_e$ unbypassed.\\
\\
\emph{Selection of $R_1$ and $R_2$~:}\\
Assume $s$ = 10
\begin{equation}
s = \frac{1 + h_{fe,max}}{1 + \frac{h_{fe,max}~\ast~R_e}{R_b + R_e}}
\end{equation}
$R_b$ is found.
\begin{equation}
V_{r2} = V_{be} + I_{CQ}\ast R_e
\end{equation}
\begin{equation}
V_{r1} = V_{CC} - V_{r2}
\end{equation}
Assume $V_{be}$ = 0.7V
\begin{equation}
\frac{R_1}{R_2} = \frac{V_{r1}}{V_{r2}} \label{poweramp1}
\end{equation}
We get $R_1$ in terms of $R_2$ and then substitute in $R_b$
\begin{equation}
R_b = R_1 \parallel R_2 = \frac{R_1~\ast~R_2}{R_1 + R_2}
\end{equation}
$R_2$ is found. Select lower standard value to make circuit independent of $\beta$. Then it is substituted in \eqref{poweramp1} to find $R_1$. Select higher standard value so that circuit draws minimum current from supply.\\
\\
\emph{Selection of output transformer~:}\\
\begin{equation}
R_{L'} = \frac{V_{CE,peak}}{I_{C,peak}}
\end{equation}
$R_{L'}$ is calculated.
\begin{equation}
R_{L'} = \frac{N_1^2}{N_2^2}~\ast~R_L
\end{equation}
$\frac{N_1}{N_2}$ is calculated.\\

\end{document}